# Spin Seebeck insulator

K. Uchida[1], J. Xiao[2,3], H. Adachi[4,5], J. Ohe[4,5], S. Takahashi[1,5], J. Ieda[4,5], T. Ota[1], Y. Kajiwara[1], H. Umezawa[6], H. Kawai[6], G. E. W. Bauer[3], S. Maekawa[4,5] and E. Saitoh[1,4,7*]

[1]*Institute for Materials Research, Tohoku University, Sendai 980-8577, Japan,*

[2]*Department of Physics and State Key Laboratory of Surface Physics, Fudan University, Shanghai 200433, China,*

[3]*Kavli Institute of NanoScience, Delft University of Technology, 2628 CJ Delft, The Netherlands,*

[4]*Advanced Science Research Center, Japan Atomic Energy Agency, Tokai 319-1195, Japan,*

[5]*CREST, Japan Science and Technology Agency, Sanbancho, Tokyo 102-0075, Japan,*

[6]*FDK Corporation, Shizuoka 431-0495, Japan,*

[7]*PRESTO, Japan Science and Technology Agency, Sanbancho, Tokyo 102-0075, Japan.*

[*]*e-mail: saitoheiji@imr.tohoku.ac.jp.*

**Thermoelectric generation is an essential function of future energy-saving technologies[1,2]. However, this generation has been an exclusive feature of electric conductors, a situation which inflicts a heavy toll on its application; a conduction electron often becomes a nuisance in thermal design of devices. Here we report electric-voltage generation from heat flowing in an insulator. We reveal that, despite the absence of conduction electrons, a magnetic insulator $LaY_2Fe_5O_{12}$ converts a heat flow into spin voltage. Attached Pt films transform this spin voltage into electric voltage by the inverse spin Hall effect[3-12]. The experimental results require us to introduce thermally activated interface spin exchange between $LaY_2Fe_5O_{12}$ and Pt. Our findings extend the range of potential materials for thermoelectric applications and provide a crucial piece of information for understanding the physics of the spin Seebeck effect[13].**

For thermoelectric power generation, we often rely on the Seebeck effect[1,2], which is the generation of electric voltage as a result of a temperature gradient (Fig. 1a). Here, electric voltage refers to the potential for electron's charge and it drives charge currents. In contrast, recent studies on spintronics[14-16] and spin caloritronics[17-19] revealed that a spin analogy for the Seebeck effect, named the spin Seebeck effect[13] (SSE), appears in magnets. The SSE stands for the generation of 'spin voltage' as a result of a temperature gradient (Fig. 1b). Spin voltage refers to the potential for electron's spins, which drives spin currents[20], i.e., net flows of spin angular momentum. The SSE has been reported in several metallic magnets, although consensus about its microscopic mechanism has not yet been reached.

The conventional Seebeck effect requires itinerant charge carriers, or conduction electrons, and therefore exists only in electric conductors (Fig. 1a). It appears natural to assume that the same should hold for the SSE. In this work, we show that a magnetic insulator $LaY_2Fe_5O_{12}$ exhibits the SSE despite the absence of conduction electrons (Fig. 1c). This result clearly indicates that thermally induced spin voltage is associated with magnetization dynamics. This new mechanism for the thermo-spin conversion also allows us to narrow down scenarios for the physical origin of the SSE in general.

The observation of the SSE in a magnetic insulator is realized by the inverse spin Hall effect[3-12] (ISHE) in Pt films. The ISHE enables to convert the spin signal generated by heat flows in the insulator into electric voltage. Figure 1d shows a schematic illustration of the sample used in the present study. It is similar to that described in ref. 13 but the magnetic metal layer is replaced by the garnet-type ferrimagnetic insulator $LaY_2Fe_5O_{12}$. The single-crystal $LaY_2Fe_5O_{12}$ (111) film with thickness of 3.9 μm was grown on a $Gd_3Ga_5O_{12}$ (111) substrate by liquid phase epitaxy. Two (and later more) 15-nm-thick Pt wires were then sputtered in an Ar atmosphere on the top of the film. Here, the surface of the $LaY_2Fe_5O_{12}$ layer has an 8×4 mm$^2$ rectangular shape. The length and width of the Pt wires are 4 mm and 0.1 mm, respectively. The resistance between the Pt wires is much greater than 10 GΩ, indicating that the wires are electrically well insulated. An in-plane external magnetic field $H$ was applied along the $x$ direction (Fig. 1d), except for the $H$-angle-dependent measurement shown in Fig. 2g. When $|H|>20$ Oe, the magnetization **M** of the $LaY_2Fe_5O_{12}$ layer is aligned along the $H$ direction (see the magnetization curve shown in Fig. 2f). A temperature gradient $\nabla T$ was applied along the $x$ direction by generating a temperature difference $\Delta T$ between the ends of the sample. Here, the temperature of one end of the sample was stabilized to $T=300$ K (Fig. 1d). Since the thermal conductivities of $LaY_2Fe_5O_{12}$ and the substrate $Gd_3Ga_5O_{12}$ are almost the same, heat currents flow uniformly along the temperature gradient. Since the localized magnetic moments in $LaY_2Fe_5O_{12}$ and conduction electrons in Pt are coupled by the interface spin-exchange interaction[21], the SSE in the $LaY_2Fe_5O_{12}$ layer should generate a spin current in the attached Pt layer with a spatial direction $\mathbf{J}_s$ ($z$ direction) and a spin-polarization vector $\boldsymbol{\sigma}$ parallel to the **M** direction ($x$ direction) (Fig. 1e). In the Pt layer, the ISHE converts this spin current into an electric field $\mathbf{E}_{ISHE}$ via the spin-orbit interaction. This electric field is generated along the $y$ direction because of the relation[5,9]

$$\mathbf{E}_{ISHE}=(\theta_{SH}\rho)\mathbf{J}_s\times\boldsymbol{\sigma} \qquad (1)$$

where $\theta_{SH}$ and $\rho$ denote the spin-Hall angle and electric resistivity of the Pt layer, respectively. Therefore, we can detect $\mathbf{E}_{ISHE}$ by measuring an electric voltage difference $V$ between the ends of the Pt wire (Figs. 1d and 1e).

Figures 2b and 2c show $V$ as a function of $\Delta T$ at $H=100$ Oe, measured when the Pt wires are attached to the lower- and higher-temperature ends of the $LaY_2Fe_5O_{12}$ layer, respectively. The magnitude of $V$ is observed to be proportional to $\Delta T$ in both Pt wires. The sign of $V$ at finite values of $\Delta T$ is clearly reversed between these wires; this distinctive behaviour of $V$ is the key feature of the SSE[13].

In Figs. 2d and 2e, we show the $H$ dependence of $V$ for various values of $\Delta T$ in the Pt wires placed at the lower- and higher-temperature ends of the $LaY_2Fe_5O_{12}$ layer, respectively. The sign of $V$ at finite values of $\Delta T$ is reversed by reversing $H$ when $|H|>20$ Oe. This sign reversal of $V$ reflects the magnetization reversal of the $LaY_2Fe_5O_{12}$ layer (cf. Figs. 2f and 2g). As also shown in Fig. 2g, this $V$ signal disappears when the direction of $H$ is perpendicular to the $x$ direction, a situation consistent with equation (1). Since $LaY_2Fe_5O_{12}$ is an insulator, the complicating thermoelectric phenomena in itinerant magnets, such as the conventional Seebeck and Nernst-Ettingshausen[22] effects, do not exist at all. In fact, the $V$ signal disappears in a $LaY_2Fe_5O_{12}$/Cu film in which the Pt wires are replaced by Cu wires with weak spin-orbit interaction (Fig. 2h), proving that the signal is generated in the Pt wires affected by the $LaY_2Fe_5O_{12}$/Pt interface. We checked that the signal disappears in a $LaY_2Fe_5O_{12}$/$SiO_2$/Pt system, where the $LaY_2Fe_5O_{12}$ and Pt layers are separated by a thin (10 nm) film of insulating $SiO_2$, as well as in a $Gd_3Ga_5O_{12}$/Pt system, where the $LaY_2Fe_5O_{12}$ layer is replaced by a paramagnetic $Gd_3Ga_5O_{12}$. These results indicate that direct contact between the magnet $LaY_2Fe_5O_{12}$ and Pt is essential for the $V$ signal generation. In the present setup, collinear orientation of the temperature gradient and magnetic field suppresses Nernst-type experimental artifacts by symmetry. An extrinsic proximity effect or induced ferromagnetism in the Pt layers is also irrelevant because of the sign change of $V$ between the ends of the $LaY_2Fe_5O_{12}$/Pt sample. Figure 2i shows that the $V$ signal disappears when the entire $LaY_2Fe_5O_{12}$/Pt sample is uniformly heated to 320 K, indicating that an in-plane temperature gradient is necessary for the observed voltage generation (note that 320 K is much lower than the Curie temperature of $LaY_2Fe_5O_{12}$ (~550 K)[23]). All results shown in Fig. 2 confirm the existence of the spin current injected into the Pt layer which is generated from the temperature gradient in the insulator $LaY_2Fe_5O_{12}$.

Next, we measured the spatial distribution of the thermally induced spin voltage. To this end, we fabricated a $LaY_2Fe_5O_{12}$/Pt sample in which nine separated Pt wires were sputtered on the top of the $LaY_2Fe_5O_{12}$ layer as depicted in Fig. 3. During the measurements, a magnetic field of 100 Oe and a uniform temperature gradient were applied along the $x$ direction. In Fig. 3, we show $V$ as a function of $x_{Pt}$, the position of the Pt wire from the centre of the $LaY_2Fe_5O_{12}$ layer along the $x$ direction, for various values of $\Delta T$. When a finite $\Delta T$ is applied, $V$ clearly increases for $x_{Pt}>0$ but decreases for $x_{Pt}<0$, similar to the behaviour of the SSE observed in metallic films[13] (see also Supplementary Information (SI)).

We now present a qualitative picture of the physics underlying this effect. Since $LaY_2Fe_5O_{12}$

is an insulator, the SSE observed here cannot be expressed in terms of thermal excitation of conduction electrons as described in our previous paper on metallic systems[13]. The SSE in insulators also cannot be explained by equilibrium "spin pumping[9,24]" as reported in ref. 21, since the average spin-pumping current from thermally fluctuating magnetic moments is exactly canceled by the thermal (Johnson-Nyquist) spin-current noise[25-27]. Therefore, the observed spin voltage requires us to introduce an unconventional non-equilibrium state between magnetic moments in LaY$_2$Fe$_5$O$_{12}$ and electrons in Pt. We can describe this excited state by effective spin-wave[28] temperature $T_F^*$ in LaY$_2$Fe$_5$O$_{12}$ and electron temperature $T_N$ in Pt equilibrated with the lattice (see SI). The $V$ signal derived from the net spin current is then proportional to $T_F^* - T_N$. Here, spin waves interact only weakly with the heat baths at both ends of the samples; they are colder than the lattice on the hot side and hotter than the lattice on the cold side[29]. Therefore, $V$ has to change sign in the centre of the sample, as observed. The close to linear dependence of the spin voltage bears witness of the extremely weak damping of the magnetization dynamics in LaY$_2$Fe$_5$O$_{12}$ that prevents a rapid equilibration of the spin waves far from the edges[21,30]. Since this spin-current generation mechanism originates from the dynamical spin coupling between LaY$_2$Fe$_5$O$_{12}$ and Pt, its physical concept is radically different from the static spin-injection model suggested previously for metallic systems[13]. The qualitative picture shown above is confirmed by the microscopic models based on scattering and linear-response theories presented in SI.

The SSE in magnetic insulators can be applied directly to the design of thermo-spin generators and, in combination with the ISHE, thermoelectric generators, allowing new approaches towards the improvement of thermoelectric generation efficiency. In general, the efficiency is improved by suppressing the energy loss due to heat conduction and Joule dissipation, which are realized respectively by reducing the thermal conductivity $\kappa$ for the sample part where heat currents flow and by reducing the electric resistivity $\rho$ for the part where charge currents flow[2]. However, in electric conductors, when $\kappa$ is dominated by electronic thermal conductivity $\kappa_e$, the Wiedemann-Franz law ($\kappa_e\rho$=constant) limits this improvement. In contrast, in the present SSE setup, $\kappa$ and $\rho$ are independent since the heat and charge currents flow in different parts of the sample: $\kappa$ stands for the thermal conductivity of the magnetic insulator and $\rho$ for the electric resistivity of the metallic wire. The SSE in insulators allows us to construct thermoelectric devices operated by an entirely new principle. The figure of merit is still small, but can be enhanced by reducing the phonon contribution to the thermal conductivity of the insulator layer and improving the interface spin-exchange efficiency and the spin-Hall angle.

**Acknowledgments**

The authors thank B. J. van Wees, K. Sato, Y. Suzuki, G. Tatara, W. Koshibae, K. M. Itoh, and M. Matoba for valuable discussions. This work was supported by a Grant-in-Aid for Scientific Research in Priority Area 'Creation and control of spin current' (19048009, 19048028), a Grant-in-Aid for Scientific Research A (21244058), the global COE for the 'Materials integration international center of education and research' all from MEXT, Japan, a Grant for Industrial Technology Research from NEDO, Japan, Fundamental Research Grants from CREST-JST, PRESTO-JST, TRF, and TUIAREO, Japan, the Dutch FOM foundation, and EC Contract IST-033749 "DynaMax".

**Author contributions**

K.U. designed the experiment, collected all of the data, and performed analysis of the data. E.S. planned and supervised the study. K.U., H.U. and H.K. fabricated the samples. T.O. and Y.K. support the experiments. J.X., H.A., J.O., S.T., J.I., G.E.W.B. and S.M. developed the explanation of the experiment. K.U., J.X., H.A., G.E.W.B. and E.S. wrote the manuscript. All authors discussed the results and commented on the manuscript.

**Additional information**

The authors declare no competing financial interests. Correspondence and requests for materials should be addressed to E.S.

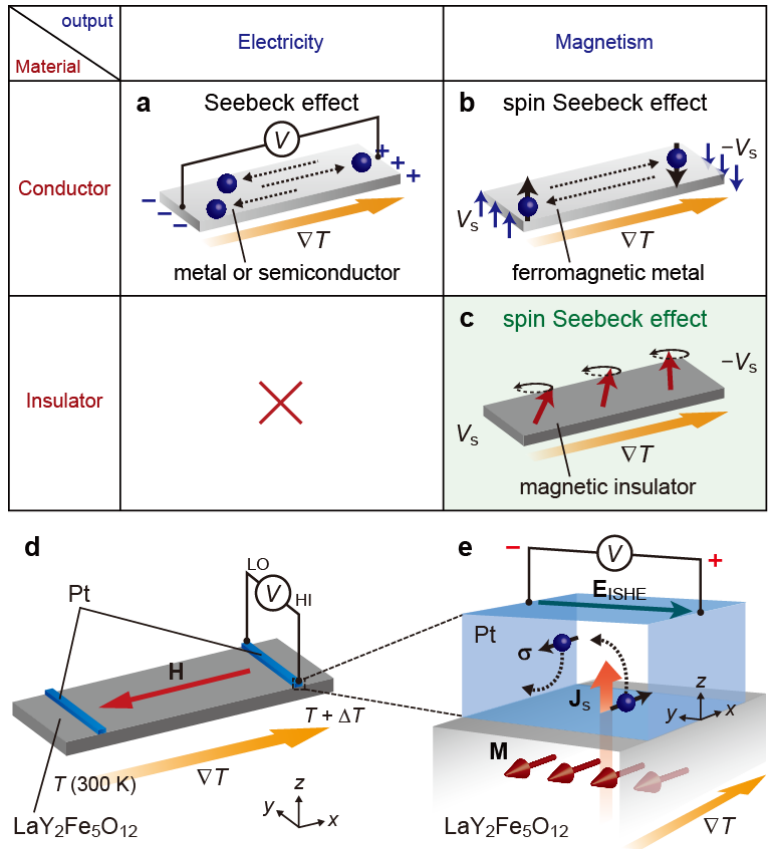

**Figure 1 | Seebeck and spin Seebeck effects. a**, A schematic illustration of the Seebeck effect. When a temperature gradient ∇$T$ is applied to a conductor, electric voltage $V$ is generated along the ∇$T$ direction. **b**,**c**, Schematic illustrations of the spin Seebeck effects. When ∇$T$ is applied to a magnet, spin voltage $V_s$ is generated. **d**, A schematic illustration of the measurement setup. The sample consists of a LaY$_2$Fe$_5$O$_{12}$ film with two Pt wires attached to the surface. An external magnetic field **H** (with magnitude $H$) and a uniform temperature gradient ∇$T$ were applied along the $x$ direction. The temperatures of the lower- and higher-temperature ends of the sample were stabilized to $T$=300 K and $T+\Delta T$, respectively, using a heater and thermocouples. **e**, A schematic illustration of the inverse spin Hall effect (ISHE) in the Pt wire and the spin current induced across the LaY$_2$Fe$_5$O$_{12}$/Pt interface. **M**, **J**$_s$, and **E**$_{ISHE}$ denote the magnetization vector of the LaY$_2$Fe$_5$O$_{12}$ layer, the spatial direction of the spin current, and the electric field generated by the ISHE in the Pt layer, respectively. The spin-polarization vector **σ** in the Pt layer is parallel with **M**.

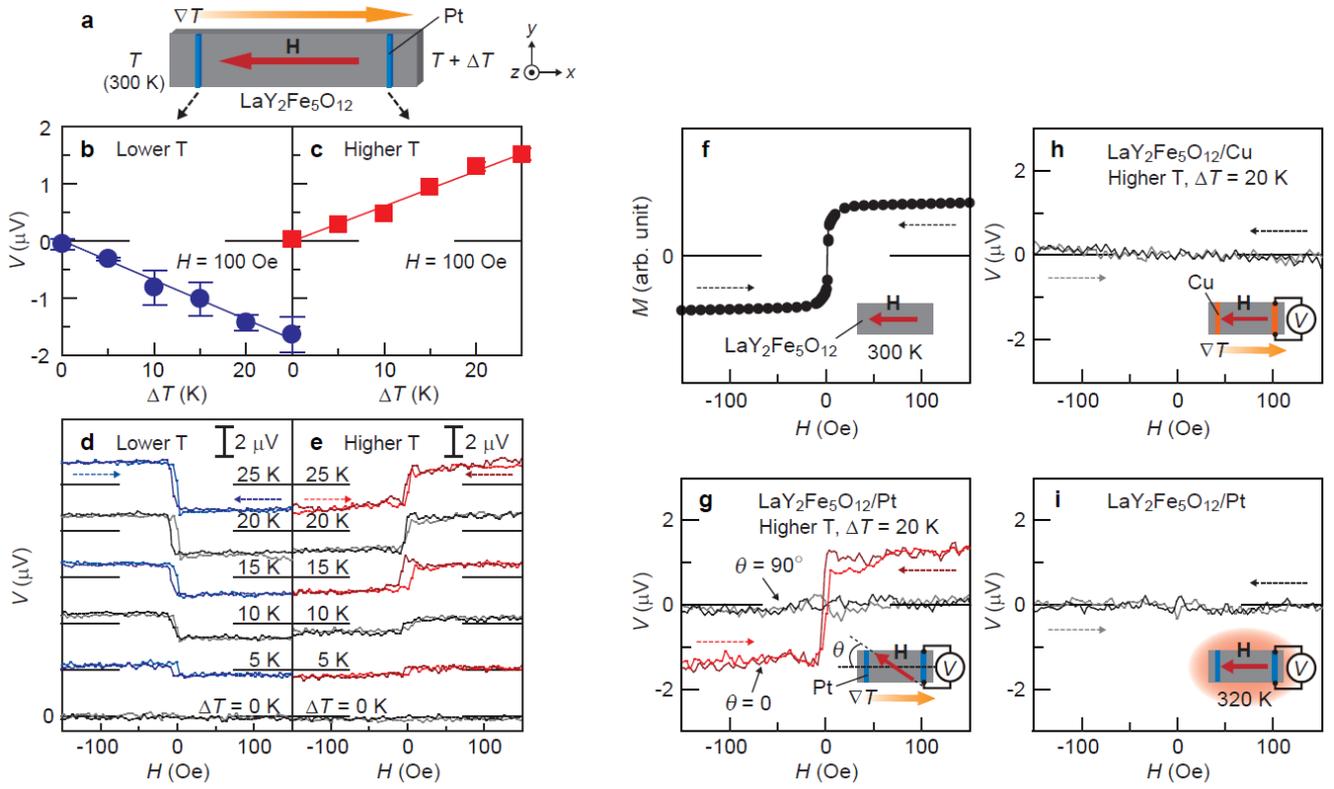

**Figure 2 | Measurements of thermal voltage generation. a**, A schematic illustration of the LaY$_2$Fe$_5$O$_{12}$/Pt sample. **b,c**, $\Delta T$ dependence of the electric voltage difference $V$ in the LaY$_2$Fe$_5$O$_{12}$/Pt sample at $H$=100 Oe, measured when the Pt wires are attached to the lower-temperature (300 K, **b**) and higher-temperature (300 K+$\Delta T$, **c**) ends of the LaY$_2$Fe$_5$O$_{12}$ layer. The error bars represent 95% confidence level. **d,e**, $H$ dependence of $V$ in the LaY$_2$Fe$_5$O$_{12}$/Pt sample for various values of $\Delta T$, measured when the Pt wires are attached to the lower-temperature (**d**) and higher-temperature (**e**) ends. **f**, Magnetization $M$ curve of the LaY$_2$Fe$_5$O$_{12}$ film at 300 K. **g**, $H$ dependence of $V$ in the LaY$_2$Fe$_5$O$_{12}$/Pt sample at $\Delta T$=20 K when the in-plane magnetic field **H** was applied at an angle $\theta$ to the $x$ direction. **h**, $H$ dependence of $V$ in a LaY$_2$Fe$_5$O$_{12}$/Cu sample at $\Delta T$=20 K when **H** was applied along the $x$ direction. The measurements shown in **g** and **h** were performed at the higher-temperature end of the samples. **i**, $H$ dependence of $V$ in the LaY$_2$Fe$_5$O$_{12}$/Pt sample when the entire sample was uniformly heated to 320 K using the same system in which the data shown in **b-e** were measured.

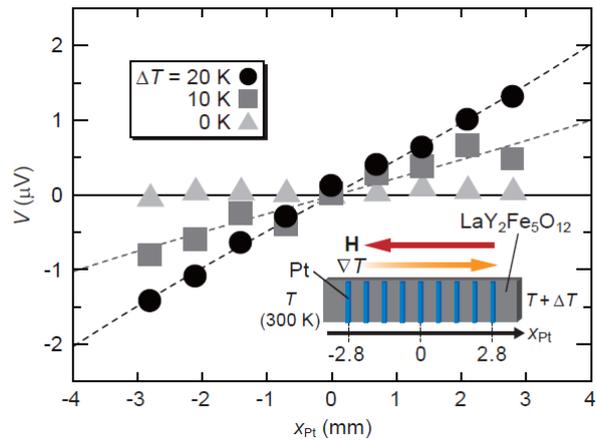

**Figure 3 | Spatial distribution.** Dependence of $V$ on $x_{Pt}$, the displacement of the Pt wire from the centre of the LaY$_2$Fe$_5$O$_{12}$ layer along the $x$ direction, in the LaY$_2$Fe$_5$O$_{12}$/Pt system for various values of $\Delta T$. In order to measure the distribution, nine Pt wires were attached to the LaY$_2$Fe$_5$O$_{12}$ layer at the intervals of 0.7 mm. Here, the middle Pt wire was attached at the centre of the LaY$_2$Fe$_5$O$_{12}$ layer ($x_{Pt}$=0 mm). The resistance between the adjacent Pt wires was much greater than 10 GΩ. The length, width, and thickness of each Pt wire are 4 mm, 0.1 mm, and 15 nm, respectively. The dashed lines represent the fitting results using our model calculation (see SI).

**A. Scattering theory**

Here we propose a mechanism for the spin Seebeck effect (SSE) observed in a magnetic insulator in terms of the effective spin-wave (magnon) temperature $T_F^*$ in the magnetic film (F, in the experiments LaY$_2$Fe$_5$O$_{12}$) and the electron temperature $T_N$ in the normal metal contact (N, in the experiments Pt).

The thermal fluctuations of the magnetization **m** at an F/N interface are excited by a thermal random magnetic field $\mathbf{h} = \sum_j \mathbf{h}^{(j)}$ ($j$ = 0, 1), which satisfies the fluctuation-dissipation theorem by the equal-position time-correlation function $\langle h_i^{(j)}(t) h_{i'}^{(j')}(t') \rangle = \left( 2k_B T^{(j)} \alpha^{(j)} / \gamma M_s V_a \right) \delta_{jj'} \delta_{ii'} \delta(t-t')$, where $\alpha^{(0)}$ is the bulk Gilbert damping parameter, $k_B$ the Boltzmann constant, $\gamma$ the gyromagnetic ratio, $M_s$ the saturation magnetization, $T^{(0)} = T_F^*$, $T^{(1)} = T_N$, and $V_a$ the magnetic coherence volume in the F layer, which depends on temperature and spin-wave stiffness constant $D$. $\alpha^{(1)} = \gamma \hbar g_r / 4\pi M_s V_a$ is the damping enhancement due to the spin pumping, where $g_r$ is the real part of the spin-mixing conductance of the interface[24].

The net thermal spin current across the F/N interface is given by the sum of the fluctuating thermal spin-pumping current $\mathbf{J}_{sp}$ from F to N proportional to $T_F^*$ and the Johnson-Nyquist spin-current noise $\mathbf{J}_{fl}$ from N to F proportional to $T_N$ (refs. 25-27):

$$\mathbf{J}_s = \mathbf{J}_{sp} + \mathbf{J}_{fl} = \frac{M_s V_a}{\gamma} \left[ \alpha^{(1)} \mathbf{m} \times \dot{\mathbf{m}} + \gamma \mathbf{m} \times \mathbf{h}^{(1)} \right] \quad (S1)$$

The DC component along the magnetization equilibrium direction ($x$ direction) reduces to

$$J_s \equiv \langle \mathbf{J}_s \rangle_x = 2\, \alpha^{(1)} k_B (T_F^* - T_N) \quad (S2)$$

The remaining task is the evaluation of the spatial profile of $T_F^* - T_N$ induced by a global temperature bias $\Delta T$ over the two ends of the F layer. Here we assume that the electron temperature of the contact, $T_N$, equals the lattice temperature in the F layer. In this situation, the temperature difference $T_F^* - T_N$ is induced at the boundaries to the heat baths, where the phonons are strongly coupled while the magnons are (almost) thermally insulated[29]. The solution of a simple heat-rate equation of the coupled magnon-phonon system yields $T_F^* - T_N = \eta \Delta T \sinh(x/\lambda_m)$, where $\eta^{-1} \approx (L/\lambda_m) \coth(L/2\lambda_m)$, $x$ is the position along the temperature-gradient direction ($x = 0$ at the centre of the F layer), and $L$ is the length of the F layer along the direction. The (squared) magnon relaxation length is described as $\lambda_m^2 \approx 3.9(Dk_BT/\hbar^2)\tau_{mm}\tau_{mp}$, where $\tau_{mm}$ ($\tau_{mp}$) is the magnon-magnon (magnon-phonon) relaxation time.

In the present LaY$_2$Fe$_5$O$_{12}$/Pt sample, the spin current shown in equation (S2) is converted into electric voltage in the Pt layer due to the inverse spin Hall effect (ISHE) leading to:

$$V_{ISHE} = \xi \Delta T \sinh(x/\lambda_m) \quad (S3)$$

where $\xi = \eta \theta_{SH} |e| k_B \rho l g_r \gamma / (\pi M_s V_a A)$, $\theta_{SH}$ is the spin-Hall angle (0.0037 for Pt[11]), $e$ the electron charge, $\rho$ the resistivity of the Pt layer, $l$ the length of the Pt layer, and $A$ the contact area. By using the parameters in Table S1, we find $\xi = 0.37$ μV/K and $\lambda_m = 4.8$ mm for LaY$_2$Fe$_5$O$_{12}$ at $T = 300$ K. These calculated values roughly agree with the experimental $\xi = 0.16$ μV/K and $\lambda_m = 6.7$ mm, which are obtained by fitting of the data shown in Fig. 3 using equation (S3) (see dashed lines in Fig. 3).

The hyperbolic sine distribution of the ISHE voltage in equation (S3) is confirmed by the temperature dependent measurement. Figure S1a shows $V/\Delta T$ as a function of $x_{Pt}$ in the LaY$_2$Fe$_5$O$_{12}$/Pt system for various values of the sample temperature $T$. When $T > 200$ K, $V$ varies almost linearly with respect to $x_{Pt}$. In contrast, below 150 K, the $x_{Pt}$ dependence of $V$ deviates from the linear function; the magnitude of $V$ decays within several millimeters from both the ends of the LaY$_2$Fe$_5$O$_{12}$ layer. In all the temperature range, the observed $V$ distribution can be fitted by equation (S3), where $\xi$ and $\lambda_m$ are fitting parameters (see solid curves in Fig. S1a). These results support our model, although the precise temperature dependence of the parameters remains to be understood.

**Table S1** | Parameters for LaY$_2$Fe$_5$O$_{12}$/Pt.

| | |
|---|---|
| $\gamma$ | $1.76 \times 10^{11}$ T$^{-1}$s$^{-1}$ |
| $4\pi M_s$ | $1.4 \times 10^5$ A/m |
| $T_c$ (Curie temp.)[23] | 550 K |
| $D$ (ref. 21) | $1.55 \times 10^{-38}$ Jm$^2$ |
| $\alpha$ | $5 \times 10^{-5}$ |
| $\tau_{mm}\tau_{mp}$ (ref. 31, 32) | $10^{-15}$ s$^2$ |
| $g_r/A$ | $10^{15}$ m$^{-2}$ |
| $V_a^{1/3}$ | 5.43 nm |
| $L$ | 8 mm |
| $l$ | 4 mm |
| $\rho$ | 0.91 μΩm |

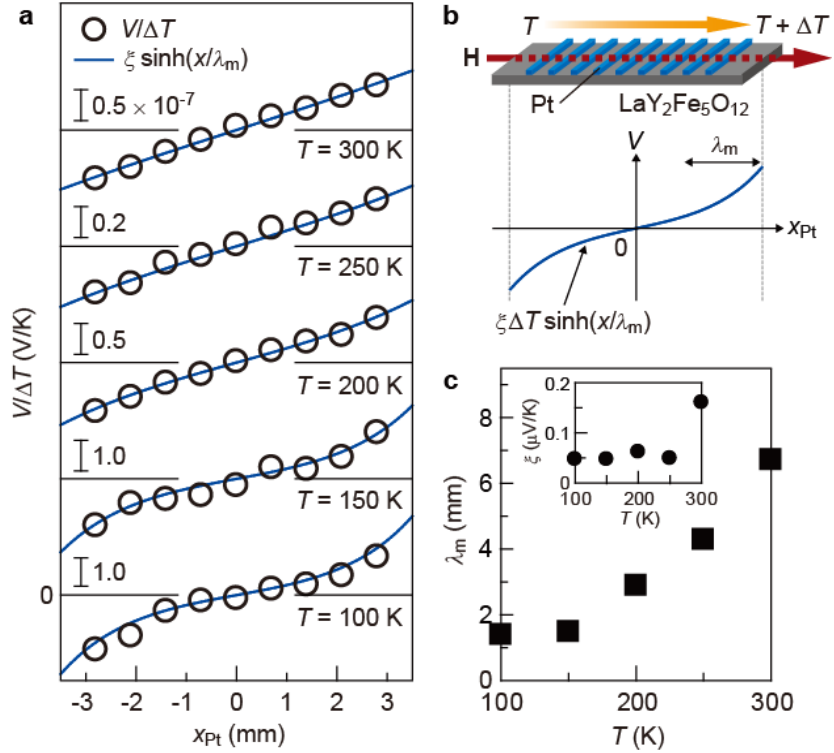

**Figure S1 | Temperature dependence. a**, Dependence of $V/\Delta T$ on $x_{Pt}$, the displacement of the Pt wire from the centre of the LaY$_2$Fe$_5$O$_{12}$ layer along the $x$ direction, in the LaY$_2$Fe$_5$O$_{12}$/Pt sample for various values of $T$ at $H = 100$ Oe (solid circles). The solid curves are the fitting results using a hyperbolic-sine function $\xi\sinh(x/\lambda_m)$, where $\xi$ and $\lambda_m$ are adjustable parameters. **b**, A schematic illustration of the LaY$_2$Fe$_5$O$_{12}$/Pt sample and the spatial distribution of $V$. **c**, $T$ dependence of $\lambda_m$. Inset to **c** shows the $T$ dependence of $\xi$.

## B. Linear-response theory

In this section, we show within the linear-response formalism that we can explain the magnitude and spatial dependence of the spin current $J_s$ in the Pt wire induced by the thermal motion of localized magnetic moments in the $LaY_2Fe_5O_{12}$ layer. The qualitative picture given in the main text is confirmed by this linear-response approach[33] as well.

We start from the Landau-Lifshitz-Gilbert equation describing the thermal motion of the localized magnetic moments in the $LaY_2Fe_5O_{12}$:

$$\partial_t \mathbf{m}(\mathbf{r},t) = \gamma[\mathbf{H}_{eff} + \mathbf{h}(\mathbf{r},t) + D\nabla^2\mathbf{m}(\mathbf{r},t)] \times \mathbf{m}(\mathbf{r},t) + \alpha \mathbf{m}(\mathbf{r},t) \times \partial_t \mathbf{m}(\mathbf{r},t) \quad (S4)$$

where $\mathbf{m} = \mathbf{M}/M_s$ is the unit magnetization normalized by the saturation magnetization $M_s$, $\alpha$ the Gilbert damping constant, $\mathbf{H}_{eff}$ the effective magnetic field, and $\mathbf{h}$ expresses the effect of thermal noise satisfying $\langle h_i(\mathbf{r},t)h_{i'}(\mathbf{r}',t')\rangle = (2k_B T(\mathbf{r})\alpha/\gamma M_s)\delta_{ii'}\delta(\mathbf{r}-\mathbf{r}')\delta(t-t')$ through the fluctuation-dissipation theorem.

The thermal motion of the localized magnetic moments in the $LaY_2Fe_5O_{12}$ affects the dynamics of the conduction-electron spin $\mathbf{s}$ in the Pt wire through the following s-d interaction at the $LaY_2Fe_5O_{12}$/Pt interface[34],

$$H_{sd} = -\Omega_{sd} S_0 \sum_{\mathbf{r}\in \text{interface}} \mathbf{s}(\mathbf{r})\cdot\mathbf{m}(\mathbf{r}) \quad (S5)$$

where $\Omega_{sd}$ is the s-d interaction strength and $S_0 = M_s a_s^3/\gamma\hbar$ is the size of the localized spin with the effective block spin volume $a_s^3$. This interaction then induces a spin current $J_s \equiv (\hbar/2)\langle\partial_t s^x\rangle$ into the Pt wire across the $LaY_2Fe_5O_{12}$/Pt interface (note that the spin quantizing direction is along the $x$ direction). Using the Heisenberg equation of motion for $s^x$, the resultant spin current $J_s$ can be calculated as an interface correlation[35] between $\mathbf{s}$ and $\mathbf{m}$:

$$J_s = \frac{\Omega_{sd} S_0}{2} \sum_{\mathbf{r}\in \text{interface}} \text{Im}\langle m^+(\mathbf{r},t)s^-(\mathbf{r},t)\rangle \quad (S6)$$

where $m^{\pm} = m^y \pm im^z$ and $s^{\pm} = s^y \pm is^z$. This interface correlation can be evaluated perturbatively in terms of the s-d interaction strength $\Omega_{sd}$.

When neglecting the influence of the temperature gradient inside of the $LaY_2Fe_5O_{12}$ layer (process $P_1$ in Fig. S2), we find that the steady-state spin current $J_s$ in the Pt wire vanishes because of the fluctuation-dissipation theorem. In contrast, when taking account of the nonzero temperature variations (process $P'_1$ and $P_3$), we obtain a nonzero spin current as

$$J_s = L_s[T(x) - T(x_0)] \tag{S7}$$

where $T(x)$ denotes the temperature at position $x$ ($x_0$ is the centre of the sample: $T(x_0) = T_2$ in Fig. S2), and $L_s = k_B \dfrac{S_0^2 \Omega_{sd} N_{int} (a/\lambda_N)^3 \chi_N \tau_{sf}}{64\sqrt{2}\pi^4 \hbar^2 \alpha (\Lambda/a_s)}$ with $\lambda_N$, $\chi_N$, $\tau_{sf}$, $a$, $N_{int}$, and $\Lambda$ being the spin diffusion length, spin susceptibility, spin relaxation time, lattice constant in the Pt wire, the number of local magnetic moments at the LaY$_2$Fe$_5$O$_{12}$/Pt interface, and the size of a temperature domain along the temperature gradient, respectively. Note that the spin current at the centre of the sample vanishes because the two relevant processes (P$_2$ and P'$_2$) cancel out.

The spin current discussed here has the same linear profile as that observed in our experiment (see Fig. 3 in the main part of the article), on the basis that the inverse spin-Hall voltage due to the SSE is proportional to the spin current induced in the Pt wire, i.e., $V_{ISHE} = \theta_{SH} J_s (2|e|/\hbar)(\rho/w)$ with $w$ being the width of the Pt wire. Note that the observed millimeter-scale spatial distribution is consistent with the recent experiment on Y$_3$Fe$_5$O$_{12}$ (ref. 30). Using $\lambda_N$ = 7 nm, $a$ = 0.2 nm, $\theta_{SH}$ = 0.0037, $\rho$ = 0.91 μΩm, $\chi_N$ = 1 × 10$^{-6}$ cm$^3$/g, $\tau_{sf}$ = 1 ps, $a_s$ = 1.23 nm, and the interface s-d coupling extracted from the independent experiment, $\Omega_{sd}$ ~ 10 meV (ref. 21), the spin current derived in equation (S7) explains the magnitude of the voltage signal $V_{ISHE}/\Delta T$ ~ 0.1 μV/K observed in our experiment.

Finally, we point out that the concept to explain the present experiment is quite different from that invoked for the previous Ni$_{81}$Fe$_{19}$-based SSE experiment. All previous models dealt with the effects of static magnetic moments in local equilibrium with the conduction electrons[36]. In contrast, here we demonstrate that the SSE in a magnetic insulator can be understood only by considering *dynamical* magnetic moments, i.e., spin waves or magnons. The finding of this unconventional mechanism suggests that magnons play a key role also in the SSE of metallic magnets. Therefore, the observation of the magnon-driven SSE in the insulator LaY$_2$Fe$_5$O$_{12}$ and its theoretical formulation have possibly larger ramifications for the physics of metallic magnets.

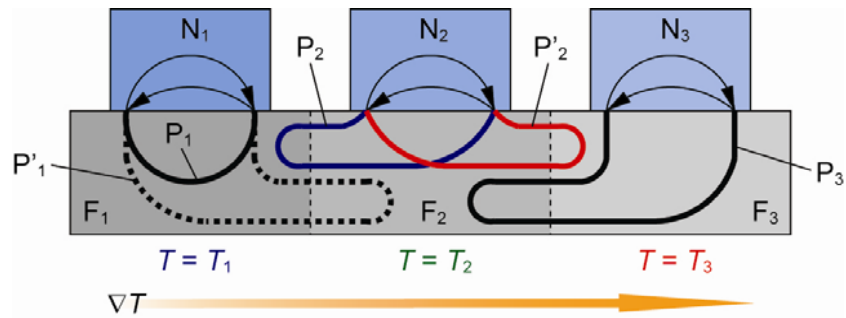

**Figure S2 | Linear-response calculation of SSE.** Feynman diagrams for calculating the spin current flowing across the interface between a magnetic insulator (F, in the experiments $LaY_2Fe_5O_{12}$) and nomal metals (N, in the experiments Pt). Here, the system is divided into three temperature domains ($N_1/F_1$, $N_2/F_2$, $N_3/F_3$) with their temperatures $T_1$, $T_2$, $T_3$. The thin solid lines with arrows (bold lines without arrows) represent electron propagators (magnon propagators).

**Additional References**